# THE INITIAL RESULTS OF RESEARCH ON TWO-STEP CASCADES IN THE DALAT RESEARCH REACTOR

Nguyen Xuan Hai[a], Pham Dinh Khang[b], Vuong Huu Tan[a], Ho Huu Thang[a],
A.M. Sukhovoj[c], V.A. Khitrov[c]

[a]*Vietnam Atomic Energy Commision, Vietnam*
[b]*Hanoi University of Science, Vietnam*
[c]*Frank Laboratory of Neutron Physics, Joint Institute for Nuclear Research, 141980 Dubna, Russia*

**Abstract:** *By the financial support of Vietnam Atomic Energy Commission (VAEC) and kind cooperation of Frank Laboratory, in the year of 2005 a measure system based on summation of amplitude pulses (SACP) was established on the tangential channel of Dalat Research Reactor. After a serial of testing, the measure system was explored. In this, we would like to show the initial results were gotten with $^{36}Cl$ isotope.*

## 1. Introduction

The nuclear parameters obtained from intensities of two-step cascades has considerably higher reliability than that obtained within known methods due to unsuccessful relation between the experimental spectra and desired parameters of the gamma-decay process. For excited levels below 2 MeV their detail spectroscopic information were known very well from investigations of (n, ), (n,e), (d,p),... reactions. However, for higher excited levels, the information is not enough because of low intensity of transitions and bad resolution of detectors with increasing energy of transitions [1].

The two-step cascades analysis was developed in Dubna without using any the model notions of level density p and radiative strength function k. It showed that the "step-like" structure in level density and corresponding deviations of k from the simple model dependencies. So, in the first time, the possibility of the "step-like" structures in the level density and corresponding thermodynamic characteristics of nucleus was pointed out in [2].

The method of summation of amplitudes of coinciding pulses from two HPGe detectors performed at the Laboratory of Neutron Physics (JINR) allows obtaining the spectroscopic information at E* 4 MeV excitation energies. It was used to investigate the rare-earth nuclei on neutron beam from the IBR-30 pulse reactor. On the tangential beam tube of Dalat research reactor using silicon filter, we received the thermal neutron flux $3.5 \times 10^6$ n.cm$^{-2}$.s$^{-1}$, cadmium ratio 78 and gamma dose 0.22 R/hr. That is reason why we planned to investigate (n,2 ) reaction on this thermal neutron beam, especially, with the collaboration of JINR. Besides, it is possible to carry out this investigation on the other filtered neutron beams received at the Dalat research reactor.

At present time, a SACP spectroscopy system was installed and the first experiments were carried on the tangential channel of Nuclear Research Institute (NRI). The two-step cascade (TSC) data of $^{36}Cl$ was showed the success of the installing process of measurements system.

## 2. The experimental

The facilities for TSC measurement was built up at the tangential horizontal channel of the 500kW light-water research reactor in Dalat. It consists of a neutron beam guide, a detector system with associated shielding and electronics.

Two semiconductor germanium detectors are used to detect the gamma rays following thermal neutron capture. Due to the shape of the beam of thermal neutrons the detectors are located in the left and in the right target. The distances between a target and the cylindrical surfaces of the germanium crystals are about 2.8 cm.

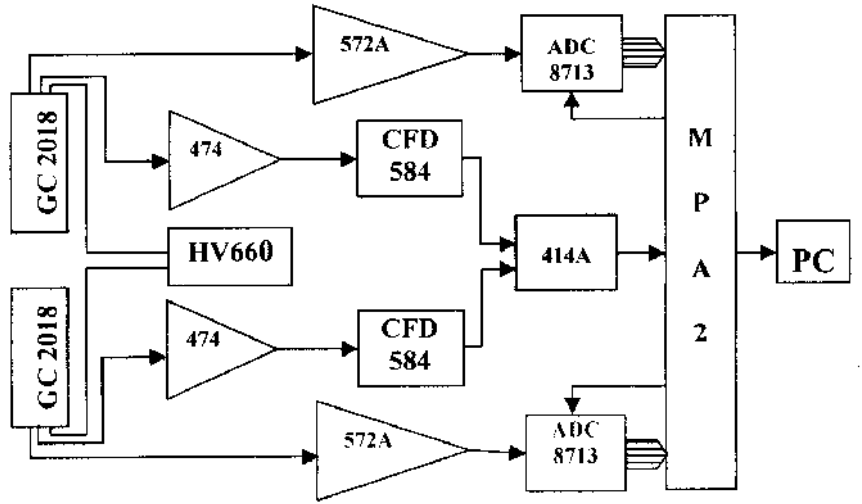

*Fig 1. The block diagram of spectroscopy system.*

The block diagram of electronic system which is based on the fast/slow coincidence system is described in Fig 1. It includes two spectroscopic amplifiers, two timing discriminators and a shaper, which forms logical pulses, make it possible for us to discard useless coincidence events with too small sums of energies. The digital outputs from two ADC's are sent to an interface and connecting with PC computer through USB ports.

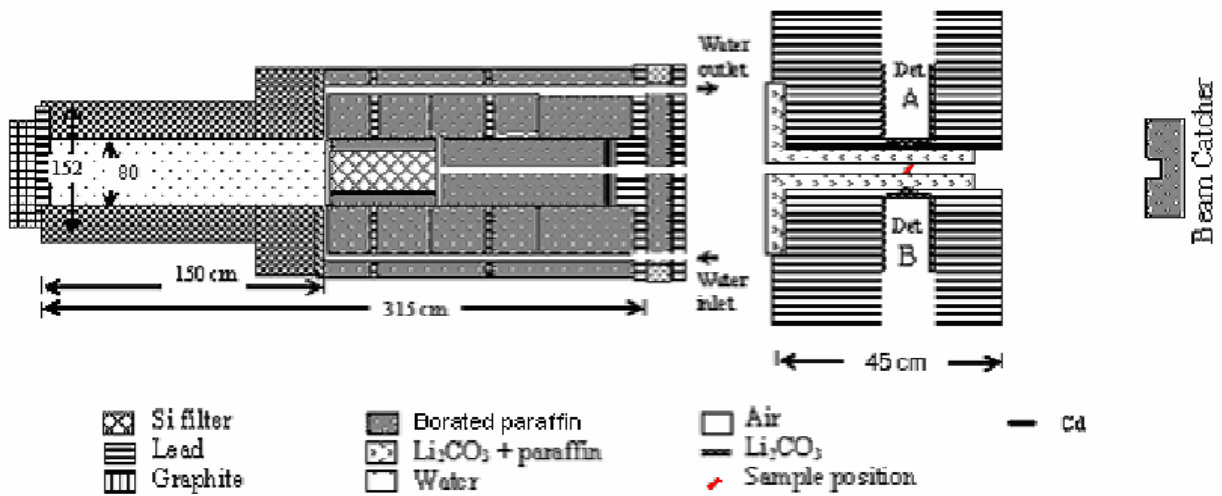

*Fig. 2. The structure of the tangential neutron channel and shielding blocks.*

To test and verify the facility and approach to data analysis, the experimental with $^{60}$Co and $^{22}$Na radioactive sources was performed in off-line situation and the experimental with the $^{35}$Cl target was performed in on-line situation. Using a NH$_4$Cl target, a nearly 100 h measurement has been performed. TSC spectra have been used to test applicability of facility and method. The experiment integrated TSC intensities have been normalized by means of the intense two-step cascade of $^{35}$Cl nucleus after thermal neutron capturing state.

To perform the symmetrization of TSC spectra one needs to know a full-energy-peak efficiency of both detectors as a function of -ray energy. The efficiency calibration was accomplished by using known emission probabilities from $^{35}$Cl(n, )$^{36}$Cl reaction in high energy region and standard radioactive multi-source for below 0.6 MeV regions. Once detectors are calibrated each channel of all TSC spectra is divided by the factor s(E )

$$s(E_\gamma) = \varepsilon_1(E_\gamma)\varepsilon_2(B_n - E_f - E_\gamma) \tag{1}$$

where $B_n$ is neutron binding energy and $E_f$ is energy of the TSC terminal state, $_1$ and $_2$ are fullenergy-peak efficiencies of the detector for which TSC spectra are accumulated (spectrumdisplaying one) and the other detector, respectively, and E is an energy deposited in the spectrumdisplaying detector.

Photon strength function $k = \Gamma_{\lambda i}/(E_\gamma^3 \times A^{2/3} \times D_\lambda)$ and level density determine the total radiative width of the compound state and cascade intensity I , obtained in the following way:

$$\Gamma_\lambda = \langle\Gamma_{\lambda i}\rangle \times m_{\lambda i} \tag{2}$$

$$I_{\gamma\gamma} = \sum_{\lambda,f}\sum_i \frac{\Gamma_{\lambda i}}{\Gamma_\lambda}\frac{\Gamma_{\lambda f}}{\Gamma i} = \sum_{\lambda,f}\frac{\Gamma_{\lambda i}}{\langle\Gamma_{\lambda i}\rangle m_{\lambda i}}n_{\lambda i}\frac{\Gamma_{if}}{\langle\Gamma_{if}\rangle m_{if}} \tag{3}$$

Here $_i$, is the partial width of -transition with energy E , A is the nucleus mass, and D is the spacing between decaying levels . The values of the total and partial gamma-widths are set for the compound state and the cascade's intermediate level i, respectively; *m* is the total number of the excited levels, and *n* is the number of excited levels below given level or *i* in the energy interval of the average cascade intensity determination.

The TSC spectra were saved in the hard disk, and processed off-line. We have developed a program called Cascad 1.0 to process TSC spectra. As the same time, the TSC spectra were processed by the program that was written by group of Sukhovoj A.M - JINR Dubna.

Bellow is some initial results that we get from the SACP spectrometer in the Dalat Research Reactor.

### 3. The results

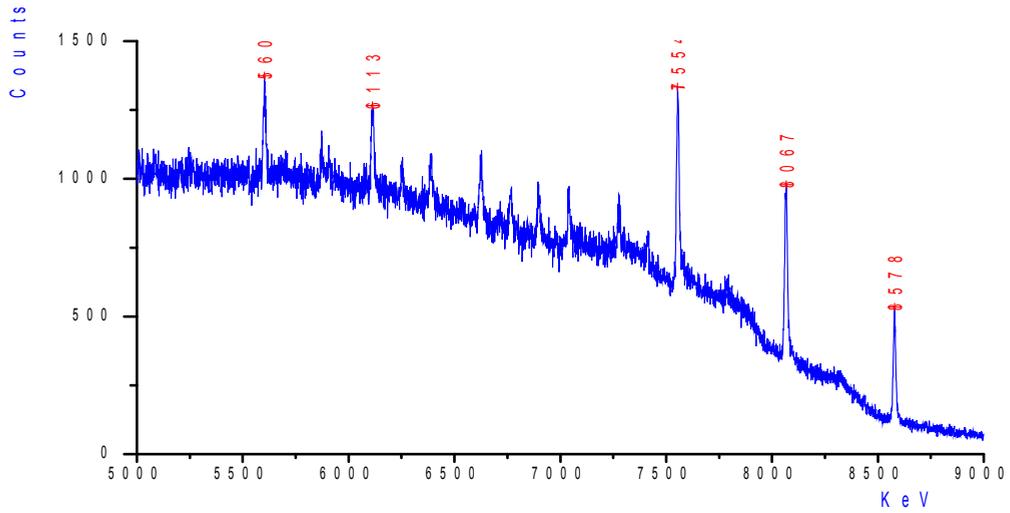

*Fig. 3. Part of the sum spectrum from $^{35}Cl(n,\gamma)^{36}Cl$ reaction with thermal neutron.*

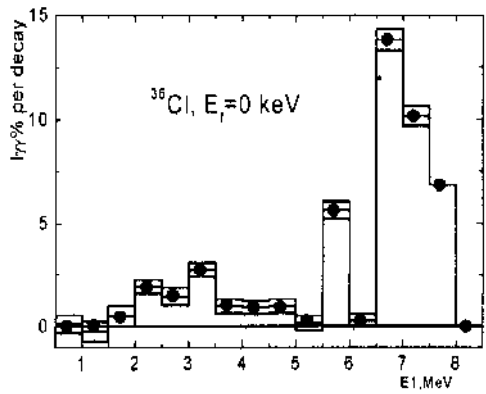

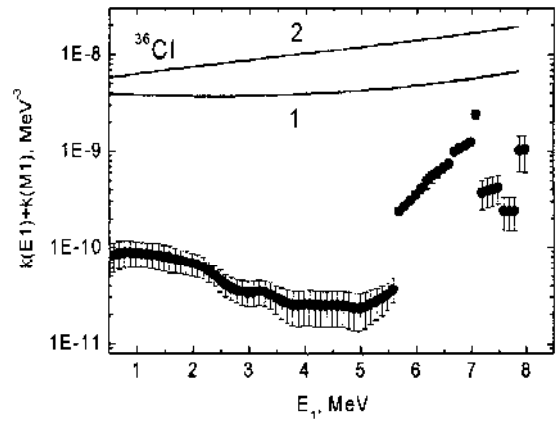

*Fig.4. Histogram is the experiment data with statistical errors, dashed points are the best fit data.*

*Fig.5. The sum of the probable radiative strength functions of E1 and M1 transitions (with estimated errors). The line 1, 2 represent predictions of the models [7] and [8].*

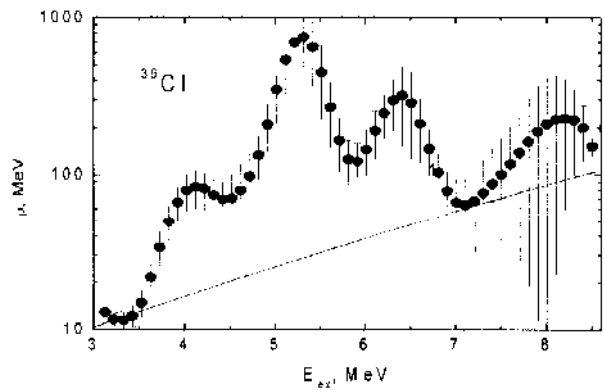

*Fig.6. The level density for $0^- \leq J^\pi \leq 2^\pm$ with their dispersion (dots with bars). Curve represents predictions of the model [6] respectively.*

*Table 1. Some gamma cascades were observed from $^{35}Cl(n,\gamma)^{36}Cl$ reaction with thermal neutron.*

| Measurement values | | | XCI 6/18/01 3 | | | I - |
|---|---|---|---|---|---|---|
| E | Up level | Low level | E | Up level | Low level | |
| 787 | 1953 | 1164 | 786.30 | 1951.20 | 1164.89 | 10.520 |
| 1164 | 1953 | 787 | 1162.78 | 1951.20 | 788.44 | 2.290 |
| 1370 | 3332 | 1959 | 1372.86 | 3332.32 | 1959.41 | 0.384 |
| 1959 | 1959 | 0.00 | 1959.36 | 1959.41 | 0.00 | 12.560 |
| 1164 | 1164 | 0.00 | 1164.87 | 1164.89 | 0.00 | 27.20 |
| 517 | 517 | 0.00 | 517.08 | 2468.28 | 1951.20 | 24.300 |
| 1951 | 2466 | 517 | 1951.14 | 1951.20 | 0.00 | 19.390 |
| 789 | 789 | 0.00 | 788.43 | 788.44 | 0.00 | 16.320 |
| 1164.59 | 1164.59 | 0.00 | 1164.87 | 1164.89 | 0.00 | 27.20 |
| 1601.5 | 1601.5 | 0.00 | 1601.08 | 1601.12 | 0.00 | 3.484 |
| 1958.5 | 1958.5 | 0.00 | 1959.36 | 1959.41 | 0.00 | 12.560 |
| 2864.3 | 2864.3 | 0.00 | 2863.82 | 2863.96 | 0.00 | 5.770 |
| 7412.6 | 8579 | 1165 | 7413.95 | 8579.70 | 1164.89 | 10.520 |
| 6979 | 8579 | 1603 | 6977.85 | 8579.70 | 1601.12 | 2.290 |
| 3063 | 8579 | 5518 | 3061.86 | 8579.70 | 5517.76 | 3.521 |
| 5518 | 5518 | 0.00 | 5517.2 | 5517.76 | 0.00 | 1.689 |
| 6621 | 8579 | 1958 | 6619.64 | 8579.70 | 1959.41 | 7.830 |
| 5716 | 8579 | 2864 | 5715.19 | 8579.70 | 2863.96 | 5.310 |
| 788.2 | 788.2 | 0.00 | 788.43 | 788.44 | 0.00 | 16.32 |
| 1950.19 | 1950.19 | 0.00 | 1951.14 | 1951.20 | 0.00 | 19.39 |
| 6628.89 | 8579 | 1950.19 | 6627.75 | 8579.70 | 1951.20 | 4.690 |
| 7791.79 | 8579 | 788.2 | 7790.32 | 8579.70 | 788.44 | 8.310 |

## 4. Analysis

Data of Fig. 5 and 6 are obtained within the framework assumptions and independences of the radiative strength functions of the primary and secondary gamma-transitions of one and the same energy and multipolarity from excitation energy of nucleus. This condition is not satisfied at least lower than 0.5Bn [11] in any nucleus. In the light nuclei, as shown into [12], this brings to the significant systematic error in determination of $\rho$ and $k$. It is concrete to the significant overestimation of $\rho$ and the underestimation $-k$. In all likelihood, this situation is manifested in $^{36}Cl$ so strongly, as in earlier studied $^{28}Al$. The insufficiency of the accumulated statistics did not make it possible to carry out the analysis [11] in the nucleus in question.

## 5. Conclusion

The data results are not much but it was showed the explanation working was done well. We hope that in the future the spectroscopy system will give more research information in the spectroscopic field and about nuclear properties in region $E_{ex}>2$-3 MeV.

We would like to thank Nuclear Research Institute and Hanoi University of Science and Frank Laboratory of Neutron Physics of Joint Institute for Nuclear Research of Dubna. The work was supported by VAEC.